\documentclass[fdp,fleqn]{w-art}
\usepackage{times}
\usepackage{w-thm}
\def\be{\begin{equation}}
\def\ee{\end{equation}}
\def\bea{\begin{eqnarray}}
\def\eea{\end{eqnarray}}
\def\barj{\bar \jmath}
\def\bari{\bar \imath}
\usepackage[]{graphicx}
\begin{document}
\DOIsuffix{theDOIsuffix}
\Volume{55}
\Issue{1}
\Month{01}
\Year{2007}
\pagespan{1}{}



\title[Constraints from F and D susy breaking]{Constraints from F and D 
supersymmetry breaking in general supergravity theories}


\author[M. Gomez-Reino]{Marta Gomez-Reino\inst{1,}%
  \footnote{Corresponding author\quad E-mail:~\textsf{marta.gomez-reino.perez@cern.ch}, 
            Phone: +0041227674523, 
            Fax: +0041227673850}}
\address[\inst{1}]{Theory Division, Physics Department, CERN, CH-1211 Geneva 23, Switzerland.}
\author[C.A. Scrucca]{Claudio A. Scrucca\inst{2,}\footnote{E-mail:~\textsf{claudio.scrucca@epfl.ch}.}}
\address[\inst{2}]{Inst. de Th. des Ph\'en. Phys.,
Ecole Polytechnique F\'ed\'erale de Lausanne, \mbox{CH-1015 Lausanne, Switzerland}}
\begin{abstract}

We study the conditions under which a generic supergravity model involving chiral and 
vector multiplets can admit vacua with spontaneously broken 
supersymmetry and realistic cosmological constant. We find that the existence 
of such viable vacua implies some constraints involving the curvature tensor of the scalar 
geometry and the charge and mass 
matrices of the vector fields, and also that the vector of $F$ and $D$ auxiliary fields defining 
the Goldstino 
direction is constrained to lie within a certain domain. We illustrate the relevance of these results 
through some examples and also discuss the implications of our general results on the 
dynamics of moduli fields in string models. This contribution is based on \cite{yo1,yo2,yo3}.

\end{abstract}
\maketitle                   

\section{Introduction}

Recently, substantial progress has been achieved in the search of non-supersymmetric
Minkowski/dS vacua in the context of string/M-theory compactifications. This was mainly related 
to the understanding of the superpotentials generated by background fluxes \cite{GVW} and by 
non-perturbative effects like gaugino condensation \cite{GC}, which generate a potential for 
the moduli fields coming from the compactification and have suggested new 
interesting possibilities for model building, like in particular those proposed in refs.~\cite{GKP,KKLT}. 
From a phenomenological point of view, this type 
of models must however posses some characteristics in order to be viable: supersymmetry must be 
broken, the cosmological constant should be tiny, and all the moduli fields should be stabilized. In 
the low energy effective theory all these crucial 
features are controlled by a single quantity, the four-dimensional scalar potential, which gives 
information on the dynamics of the moduli fields, on how supersymmetry is broken and on
the value of the cosmological constant. The characterization of the conditions under which a
supersymmetry-breaking stationary point of the scalar potential satisfies simultaneously the flatness
condition (vanishing of the cosmological constant) and the stability condition (the stationary point is
indeed a minimum) is therefore very relevant in the search of phenomenologically viable string models. 
In this note we review the techniques presented in \cite{yo1,yo2,yo3} to study the 
possibility of getting this type of vacua in the context of general supergravity 
theories in which both chiral and vector multiplets participate to supersymmetry breaking.

\section{Viable supersymmetry breaking vacua}

The goal of this section is to find conditions for the existence of
non-supersymmetric extrema of the scalar potential of general supergravity theories fulfilling two basic
properties: i) they are locally stable and ii) they lead to a negligible cosmological constant. We will 
first study this issue for theories with only chiral multiplets and then when 
also vector multiplets are present.

\subsection{Constraints for chiral theories}

The Lagrangian of the most general supergravity theory with $n$ chiral superfields is entirely defined
by a single arbitrary real function $G$ depending on the corresponding chiral superfields $\Phi_i$ and
their conjugates $\bar \Phi_i$, as well as on its derivatives \cite{sugra}. The function $G$ can be 
written in terms of a real K\"ahler potential $K$ and a holomorphic superpotential $W$ in the following 
way\footnote{We will use the standard notation in which subindices $i$, $\bar \jmath$ mean 
derivatives with respect to $\Phi^i$, $\bar \Phi^{j}$ and Planck units, $M_P=1$.}:
\be
G(\Phi_i,\bar \Phi_i) = K(\Phi_i,\bar\Phi_i) + \log W(\Phi_i) +\log \bar W(\bar\Phi_i) \,.
\ee
The quantities $K$ and $W$ are however defined only up to K\"ahler transformations acting as
$K \to K + f + \bar f$ and $W \to e^{-f} W $, $f$ being an arbitrary holomorphic function of the superfields,
which leave the function $G$ invariant. The scalar components of the chiral multiplets span an 
$n$-dimensional K\"ahler manifold whose metric is given by $G_{i\bar \jmath}$, which 
can be used to lower and raise indices. 

The 4D scalar potential of this theory takes the following simple form:
\be
V = e^G \Big(G^k G_k - 3 \Big) \,.
\label{genpot}
\ee
The auxiliary fields of the chiral multiplets are fixed by the Lagrangian through the equations 
of motion, and are given by $F_i = - \, e^{G/2}\, G_{i}$ where $e^{G/2}=m_{3/2}$ 
is the mass of the gravitino. Whenever $F_i\neq0$ at the vacuum, supersymmetry is spontaneously 
broken and the direction given by the $G_i$'s defines the direction of the Goldstino eaten 
by the gravitino in the process of supersymmetry breaking.

In order to find local non-supersymmetric minima of the potential (\ref{genpot}) with small non-negative 
cosmological constant, one should proceed as follows: First impose the condition that the 
cosmological constant is negligible and fix $V =0$. This flatness condition implies that:
\be
\label{fc}
G^k G_k = 3 \,.
\ee
Then look for stationary points of the potential where the flatness condition is satisfied. 
This implies:
\be\label{sc}
G_i + G^k \nabla_i G_k = 0 \,,
\ee
where by $\nabla_i G_k=G_{ik}-\Gamma^n_{ik}G_n$ we denote the covariant derivative with 
respect to the K\"ahler metric.

Finally, make sure that the matrix of second derivatives of the potential,
\begin{equation}
m^{2} = \left(
\begin{matrix}
m^2_{i \bar\jmath} &  m^2_{ij} \smallskip \cr
m^2_{\bar i \bar \jmath} & m^2_{\bar \imath j}
\end{matrix}
\right) \,,
\label{VIJ}
\end{equation}
is positive definite. This matrix has two different $n$-dimensional blocks, $m^2_{i \bar \jmath} = 
\nabla_i \nabla_{\bar \jmath} V$ and $m^2_{i j} = \nabla_i \nabla_j V$, and after a straightforward computation these are found to be given by the following expressions:
\begin{eqnarray}\label{vij}
\begin{array}{lll}
m^2_{i \bar \jmath} \!\!\!&=&\!\!\! e^G\Big(G_{i \bar \jmath} + \nabla_i G_k \nabla_{\bar
\jmath} G^k - R_{i \bar \jmath p \bar q} G^p G^{\bar q}\Big) \,, \smallskip\ \\
m^2_{i j} \!\!\!&=&\!\!\! e^G\Big(\displaystyle{\nabla_i G_j + \nabla_j G_i + \frac 12 G^k 
\big\{\nabla_i,\nabla_j \big\}G_k}\Big) \,, \\
\end{array}
\end{eqnarray}
where $R_{i \bar \jmath p \bar q}$ denotes the Riemann tensor with respect to the K\"ahler metric. 
The conditions under which this $2n$-dimensional matrix (\ref{VIJ}) is positive definite are 
complicated to work 
out in full generality, the only way being the study of the behaviour of the $2n$ eigenvalues. 
Nevertheless a necessary condition for this matrix to be positive definite can be encoded 
in the condition that the quadratic form $m^2_{i \bar \jmath} z^i {\bar z}^{\bar \jmath}$ is positive
for any choice of non-null complex vector $z^i$. Our strategy will be then to look for a special  
vector $z^i$ which leads to a simple constraint. 

In this case there is only one special direction in field space, that is the direction given by $z^i = G^i$. 
Indeed projecting in that direction we find the following simple expression:
\be\label{eq}
m^2_{i \bar \jmath} G^i G^{\bar \jmath} = 6 - R_{i \bar \jmath p \bar q}\, G^i G^{\bar \jmath} G^p 
G^{\bar q}
\,.
\ee
This quantity must be positive if we want the matrix (\ref{VIJ}) to be positive definite
\footnote{Actually, as emphasized in \cite{yo4}, the Goldstino multiplet cannot receive any supersymmetric 
mass contribution from $W$, since in the limit of rigid supersymmetry its fermionic 
component must be massless. This means that, in order to study metastability, it is enough to study 
the projection of the diagonal block $m^2_{i \bar \jmath} $ of the mass matrix along the 
Goldstino direction $G^i$, as the rest of the projections can be given a mass with the help of 
the superpotential.}. Using the 
rescaled variables $f^i =  - \frac{1}{\sqrt{3}} G^i$ the conditions for the existence of 
non-supersymmetric flat minima can then be written as:
\bea\label{fs-1}
\left\{\!\!
\begin{array}{l}
G_{i\,\bar \jmath}f^if^{\bar \jmath} = 1\,,\\
R_{i \,\bar \jmath \,p \,\bar q}\, f^i f^{\bar \jmath} f^p f^{\bar q}< \displaystyle{\frac23}\,.
\end{array}\right.
\eea
The first condition, the flatness condition, fixes the amount of supersymmetry breaking
whereas the second condition, the stability condition, requires the existence of directions with 
K\"ahler curvature less than {$2/3$} and constraints the direction of supersymmetry breaking 
to be sufficiently aligned with it.

\subsection{Constraints for gauge invariant theories}

It can happen that the supergravity theory with $n$ chiral multiplets $\Phi^i$ we just described 
has a group of some number $m$ of global symmetries, compatibly with supersymmetry. In this 
subsection we consider the possibility of gauging such isometries with the introduction of 
vector multiplets. The corresponding supergravity theory will then include in addition to the $n$ chiral 
multiplets $\Phi^i$, $m$ vector multiplets $V^a$. 

The two-derivative Lagrangian is specified in this case by a 
real K\"ahler function $G(\Phi^k,\bar\Phi^{k},V^a)$, determining in particular the scalar geometry, 
$m$ holomorphic Killing vectors $X_a^i(\Phi^k)$, generating the isometries that are gauged, and an 
$m$ by $m$ matrix of holomorphic gauge kinetic functions $H_{ab}(\Phi^k)$, defining the 
gauge couplings\footnote{The gauge kinetic function $H_{ab}$ must have an appropriate 
behavior under gauge transformations, in such a way as to cancel possible gauge anomalies 
$Q_{abc}$. Actually, the part 
$h_{ab}={\rm Re}\,H_{ab}$ defines a metric for the gauge fields and must be gauge invariant. 
On the other hand ${\rm Im}\,H_{ab}$ must have a variation that matches the coefficient of 
$Q_{abc}$, namely $X_a^i h_{bc i} =  \frac i2 \, Q_{abc}$.}. In this case the minimal coupling between chiral and vector multiplets turn ordinary 
derivatives into covariant derivatives, and induces a new contribution to the scalar potential 
coming from the vector auxiliary fields $D^a$, in addition to the standard one coming from the 
chiral auxiliary fields $F^i$. The 4D scalar potential takes the form:
\be
\label{genpot2}
V =  e^G\Big( g^{i \barj}\, G_i G_{\bar \jmath}-3\Big) + \frac 12 h^{ab} D_a D_b\,.
\ee
The auxiliary fields are fixed from the Lagrangian through the equations of motion  to be:
\bea
&& F_i = - m_{3/2}\, G_i \,, \label{F} \\[1mm]
&& D_a = -G_a =  i \, X_a^i \, G_i = - i \, X_a^{\bari} \, G_{\bari} \,, \label{D}
\eea
where to get the relations in (\ref{D}) one should also use gauge invariance of the action.

Now in order to find local non-supersymmetric minima of the potential (\ref{genpot2}) 
with small non-negative cosmological constant, we will proceed as in the previous subsection. 
First we will impose the condition that the cosmological constant is negligible and fix $V =0$. 
This flatness condition implies that:
\be
- 3 + G^i G_i + \frac 12 \, e^{-G} D^a D_a = 0\,.
\label{flatness}
\ee
The stationarity conditions correspond now to the requirement that $\nabla_i V = 0$, and 
they are given by:
\be
G_i + G^k \nabla_i G_k + e^{-G} \Big[D^a\Big(\nabla_i - \frac 12\, G_i \Big) D_a 
+ \frac 12\,  h_{abi} D^a D^b \Big] = 0\,.
\label{stationarity}
\ee

The $2n$-dimensional mass matrix (\ref{VIJ}) for small fluctuations of the scalar fields around the 
vacuum has as before two different $n$-dimensional blocks, which can be computed as 
$m_{i \bar\jmath}^2 = \nabla_i \nabla_{\bar\jmath} V$ and $m_{i j}^2 = \nabla_i \nabla_j V$. Using 
the flatness and stationarity conditions, one finds, after a straightforward computation 
\cite{FKZ,dudasvempati}:
\bea
&&\hspace{-1.5cm}m_{i \barj}^2 = e^G \Big[g_{i \barj} - R_{i \barj p \bar q} G^p G^{\bar q} 
+  \nabla_i G_k \nabla_{\barj} G^k \Big] - \frac 12\, \Big(g_{i \barj} - G_i G_{\barj} \Big) D^a D_a
-\,2\, D^a G_{(i} \nabla_{\barj)} D_a \label{mijbar} \\
&&\hspace{-.5cm} + \Big(G_{(i} h_{ab \barj)} + h^{cd} h_{a c i} h_{b d \barj} \Big)\, D^a D^b  - 
2\, D^a h^{bc} h_{ab(i} \nabla_{\barj)} D_c
+ h^{ab} \nabla_i D_a  \nabla_{\barj} D_b + D^a \nabla_i \nabla_{\barj} D_a\,, \nonumber \\
&& \hspace{-1.5cm}m_{i j}^2 = e^G \Big[2 \, \nabla_{(i} G_{j)} + G^k \nabla_{(i} \nabla_{j)} G_k \Big] - \frac 12 \Big(\nabla_{(i} G_{j)} - G_i G_j \Big) D^a D_a+ h^{ab}\, \nabla_i D_a \nabla_j D_b
  \label{mij} \\
&&  \hspace{-.5cm}-\,2\, D^a G_{(i} \nabla_{j)}  D_a - 2\,D^a h^{bc} h_{ab(i} \nabla_{j)} D_c 
+ \Big(G_{(i} h_{abj)} + h^{cd} h_{a c i} h_{b d j} - \frac 12 h_{a b i j} \Big) D^a D^b \,.\nonumber
\eea

We want to analyze now the restrictions imposed by the requirement 
that the physical squared mass of the scalar fields are all positive. In general the theory displays a spontaneous breakdown of both supersymmetry and gauge symmetries, so in the study of the stability 
of the vacuum it is 
necessary to take appropriately into account the spontaneous breaking of gauge symmetries. 
In that process $m$ of the $2n$ scalars, the would-be Goldstone 
bosons, are absorbed by the gauge fields and get a positive mass, so we do not need to take them 
into account for the analysis of the stability. Nevertheless the would-be 
Goldstone modes correspond to flat directions of the unphysical mass matrix, 
and get their physical mass through their kinetic mixing with the gauge bosons. 
This means that positivity of the physical mass matrix implies semi-positivity of the unphysical mass 
matrix in (\ref{mijbar}), (\ref{mij}). We can use then the same strategy as before but changing the 
strictly positive condition to a semi-positive one. 

In this case there exist two types of special complex directions $z^i$ one could look at. The first is 
the direction $G^i$, which is associated with the Goldstino direction in the subspace of chiral multiplet 
fermions. Projecting into this direction one finds, after a long but straightforward computation:
\bea
m^2_{i \bar \jmath} G^i G^{\barj} \!\!\!&=&\!\!\! 
e^G \Big[6 -R_{i \barj p \bar q} \, G^i G^{\barj} G^p G^{\bar q} \Big]
+\, \Big[\!-\! 2\, D^a D_a + h^{cd} h_{a c i} h_{b d \barj} \, G^i G^{\barj} D^a D^b \Big] \label{mGG}\\
\!\!\!&\;&\!\!\! +\,e^{-G} \Big[M^2_{ab} D^a D^b \!+\! \frac 34\,Q_{abc} D^a D^b D^c 
\!-\! \frac 12 \Big(D^a D_a\Big)^2 \! \!+\! \frac 14 h_{ab}^{\;\;\;i} h_{cdi} D^a D^b D^c D^d \Big]\,, \nonumber 
\eea
where $Q_{abc}=-2iX^i_ah_{bci}$. 
The condition $m^2\hspace{-4pt}{}_{i \barj} \, G^i G^{\barj} \ge 0$ is then the generalization of the 
condition in (\ref{eq}) for theories involving only chiral multiplets. In terms of the rescaled variables:
\be
f_i=\frac{1}{\sqrt{3}}\frac{F_i}{m_{3/2}}=-\frac{1}{\sqrt{3}}G_i\,,\hspace{1.8cm}
d_a=\frac{1}{\sqrt{6}}\frac{D_a}{m_{3/2}}\,,
\ee
the flatness and stability conditions take then the following form:
\begin{eqnarray}
\hspace{-.8cm}\left\{\hspace{-4pt}
\begin{array}{l}
\displaystyle{ \,f^i f_{i} + d^a d_a} = 1\,,\\
\displaystyle{R_{i \,\bar \jmath\, p\, \bar q} \, f^i f^{\bar \jmath} f^p f^{\bar q} \le \frac23 
+ \frac23 \Big({M_{ab}^2}/{m^2_{3/2}} -2 h_{a\,b} \Big) d^a d^b
+2 h^{c\,d}h_{a\,c\,i}h_{b\,d\,\bar \jmath}f^if^{\bar \jmath}d^ad^b}\\
\displaystyle{\hspace{2.8cm}- \Big(2\, h_{a\,b} h_{c\,d}- h_{a\,b}^ih_{c\,d\,i}\Big)\, d^a d^b d^c d^d
+\sqrt{\frac 32} \frac{Q_{abc}}{m_{3/2}} d^a d^b d^c\,.}
\end{array}
\right.
\end{eqnarray}

Again we have that the flatness condition fixes the amount of supersymmetry breaking whereas 
the stability condition constrains its direction. One could also consider the 
directions $X_a^i$, which are instead associated with the Goldstone directions in the space of chiral 
multiplet scalars. Nevertheless the constraint $m^2_{i\bar \jmath}X_a^iX_a^{\bar \jmath}\ge0$ turns 
out to be more complicated and no useful condition seems to emerge from it.

\section{Analysis of the constraints}

The analysis of the flatness and stability conditions in the case where both chiral and vector multiplets 
participate to supersymmetry breaking presents an additional complication with respect to the case 
where only chiral multiplets are present, due to the fact that the auxiliary 
fields of the chiral and vector multiplets are not independent of each other. 
The rescaled auxiliary fields $f_i$ and $d_a$ are actually related in several ways.
One first relation (consequence of gauge invariance) can be read from 
eq. (\ref{D}) and is given by:
\be
d^a =  \frac{i\,X^a_i}{\sqrt{2} m_{3/2}} \,  f^i \label{DFkin} \,.
\ee
This relation is satisfied as a functional relation valid at any point of the scalar field space. 
It shows that the $d_a$ are actually linear combinations of the $f_i$. Using 
now the inequality $|a^i b_i| \le \sqrt{a^i a_i} \sqrt{b^j b_j}$ one can derive a simple bound on 
the sizes that the $d_a$ can have relative to the $f_i$:
\be
|d_a| \le \frac {1}{2} \frac {M_{aa}}{m_{3/2}} \sqrt{f^i f_i}\,.
\label{DF}
\ee
There is also a second relation between $f_i$ and $d_a$, that is instead valid 
only at the stationary points of the potential. 
It arises by considering a suitable linear combination of the stationarity conditions along the 
direction $X_a^i$, in other words, by imposing $X_a^i \nabla_i V= 0$. This relation reads 
\cite{KawamuraDFF,ChoiDFF} (see also \cite{KawaKobDFF}):
\be
i\,\nabla_i X_{a\bar \jmath} \, f^i f^{\bar \jmath} 
-  \sqrt{\frac23}m_{3/2}\Big(3 f^i f_i - 1 \Big) \,d_a - \, \frac{M^2_{ab}}{\sqrt{6} \,m_{3/2}} \, d^b 
+Q_{abc}\,d^b d^c= 0\label{DFF}\,.
\ee
These relations show that whenever the $f_i$ auxiliary fields vanish also the $d_a$ auxiliary fields 
should vanish. Therefore we can say that the {$f_i$}'s represent the basic qualitative seed 
for supersymmetry breaking whereas the {$d_a$}'s provide additional quantitative effects. 
Along this section we will address 
the problem of working out more concretely the implications of these constraints. In order to do so 
we will concentrate on the case in which the gauge kinetic function is constant and diagonal:  
$h_{ab} = g_a^{-2} \delta_{ab}$. In this case we can rescale the vector fields in such a way as to  
include a factor $g_a$ for each vector index $a$. In this way, no explicit dependence on $g_a$ is left 
in the formulas and the metric becomes just $\delta_{ab}$. Using this the flatness and stability 
conditions take the following simple form:
\begin{eqnarray}\label{fs-2}
\left\{\hspace{-4pt}
\begin{array}{l}
 f^i f_i +\sum_a d_a^2 = 1 \,,\\
R_{i\, \bar \jmath\, p\, \bar q} \, f^i f^{\bar \jmath} f^p f^{\bar q} \le \displaystyle{\frac 23} 
+ \displaystyle{\frac 43} \, \mbox{$\sum_{a}$} \Big(2\, m_{a}^2 -1 \Big)\, d_a^2 
- 2\, \mbox{$\sum_{a,b}$} d_a^2 d_b^2\,, 
\end{array}
\right.
\end{eqnarray}
where we have defined the quantity $m_{a} =  {M_{a}}/{(2\, m_{3/2})}$ 
measuring the hierarchies between scales. Denoting $v_a^i =  {\sqrt{2} X_a^i}/{M_a}$ and 
$T_{a\, i \bar \jmath} = {i\,\nabla_i X_{a\,\bar \jmath}}/{M_a}$ the relations between 
{$f^i$} and {$d^a$} read:
\begin{eqnarray}
&& d_a = i\, m_a v_a^i f_i  \hspace{.4cm}\Longrightarrow\hspace{.4cm} 
|d_a| \le m_{a} \, \sqrt{ f^i f_i}  \,,\\
&& d_a = \sqrt{\frac 32} \, \frac {m_a \,T_{a\, i \,\bar \jmath}  \, f^i f^{\bar \jmath}}
{m_{a}^2 - 1/2 + 3/2\, f^i f_i}  \label{KIn}\,.
\end{eqnarray}

\subsection{Interplay between F and D breaking effects}

In this subsection we will study the interplay between the {$F$} and {$D$} supersymmetry 
breaking effects. In order to do so it is useful to introduce the variables ${\hat f}^i =  {f^i}/
{\sqrt{1 - \sum_a d_a^2}}$. Using these variables the conditions for flatness and stability can 
be rewritten as:
\begin{eqnarray}
\left\{\hspace{-4pt}
\begin{array}{l}
\displaystyle{ {\hat f}^i {\hat f}_{i} = 1 } \,,\\
\displaystyle{R_{i\, \bar \jmath\, p\, \bar q} \, {\hat f}^i {\hat f}^{\bar \jmath} {\hat f}^p {\hat f}^{\bar q} 
\le \frac 23 \, K(d_a^2,m_a^2) }\,,
\end{array}
\right.
\end{eqnarray}
where the function $K(d_a^2,m_a^2)$ is given by:
\be
K(d_a^2,m_a^2) = 1 + 4\, \frac {\sum_a m_a^2 d_a^2 - \big(\sum_a d_a^2\big)^2 \raisebox{-5pt}{}}
{\big(1 - \sum_b d_b^2\big)^2} \,.
\ee

In the limit in which the rescaled vector auxiliary fields are small ($d_a \ll 1$) we have that 
${\hat f}^i \simeq f^i$ and therefore these variables $\hat f^i$ are the right variables to study  
the effect of vector multiplets with respect to the case where only chiral multiplets are present. Note that 
in such a limit the relation (\ref{KIn}) between F and D auxiliary fields can be written at first order as 
$d_a\,\simeq\,\sqrt{3/2}\,m_a/(1+m_a^2)\,T_{a\,i\,\bar \jmath}\,{\hat f}^i\,{\hat f}^{\bar \jmath}$. 
Using this we get:
\be
K\,\simeq\,1+6\, \mbox{$\sum_a$} \,\xi^2_a(m)\,T_{a\,i\,\bar \jmath} \,T_{a\,p\,\bar q} \,
{\hat f}^i {\hat f}^{\bar \jmath}{\hat f}^p {\hat f}^{\bar q}\,,\hspace{1cm}\xi_a(m)=\frac{m_a^2}{1+m_a^2}\,,
\ee
and we can write the flatness and stability conditions as:
\begin{eqnarray}
\left\{\hspace{-4pt}
\begin{array}{l}
\displaystyle{ {\hat f}^i {\hat f}_{i} = 1 } \,,\\
\displaystyle{\hat{R}_{i\, \bar \jmath\, p\, \bar q} \, {\hat f}^i {\hat f}^{\bar \jmath} {\hat f}^p 
{\hat f}^{\bar q} \le \frac 23 }\,,
\end{array}
\right.
\end{eqnarray}
where $\hat{R}_{i\, \bar \jmath\, p\, \bar q} = R_{i\, \bar \jmath\, p\, \bar q} - 
4\,\sum_a \xi^2_a(m)\,T_{a\,i\,(\bar \jmath} \,T_{a\,p\,\bar q)}$. This means 
that the net effect in this case is to change 
the curvature felt by the chiral multiplets. Note as well that in the case in which the mass of the 
vectors is large this is not necessarily a small effect and can compete with the 
curvature effects due to the chiral multiplets. Actually for heavy vector fields one can check that 
integrating out the vector fields modifies the K\"ahler potential of the chiral multiplets 
in a way that accounts for this shift in the K\"ahler curvature.

For larger values of {$d_a$} one can instead find an upper 
bound to {$K$} (see \cite{yo3} for details):
\be
\hspace{-.5cm}K \leq 1+6 \, \mbox{$\sum_a$} \, \xi^2_a(m) T_{a\,i\,\bar \jmath} \,T_{a\,p\,\bar q} \,
{\hat f}^i {\hat f}^{\bar \jmath}{\hat f}^p {\hat f}^{\bar q}\,,\hspace{1cm}
\xi_a(m)=\frac{m_a^2\, (1+\sum_bm_b^2)}{1+m_a^2+(m_a^2-\frac12)
\sum_bm_b^2}\,.
\ee
So in this general case we get as well that the effect of vector multiplets can be encoded into an  
effective curvature $\hat{R}_{i\, \bar \jmath\, p\, \bar q} = R_{i\, \bar \jmath\, p\, \bar q} - 
4\,\sum_a \xi^2_a(m)\,T_{a\,i\,(\bar \jmath} \,T_{a\,p\,\bar q)}$.

In this section we have derived the implications of the flatness and stability conditions taking into 
account the fact that  $f^i$ and $d^a$ are not independent variables. The strategy that we have followed  
is to use the the relation (\ref{DFkin}) to write $d_a$ in terms of $f^i$. A second possibility would be 
to use instead the relation (\ref{DFF}) to write $d^a$ in term of $f^i$ and a third one would be to 
impose only the bound (\ref{DF}) to restrict the values of the $d^a$ in terms of the values of $f^i$. 
It is clear that switching from the relation (\ref{DFkin}) to the relation (\ref{DFF}) and finally to 
the bound (\ref{DF}) represents a gradual simplification of the formulas, which is also 
accompanied by a loss of information. As a consequence, these different types of strategies will 
be tractable over an increasingly larger domain of parameters, but this will be 
accompanied by a gradual weakening of the implied constraints. A detailed 
derivation of the implications of the flatness and stability conditions when the relations (\ref{DFF}) 
and (\ref{DF}) are used can be found in \cite{yo3}.

\section{Some examples: moduli fields in string models}

In this section we will apply our results to the typical situations arising for the moduli sector of string models. The K\"ahler potential and superpotential governing the dynamics of 
these moduli fields typically have the general structure:
\be
\label{stringk}
K = - \,\mbox{$\sum_i$}\, n_i\, {\rm ln} (\Phi_i + \bar\Phi_i) + \dots\,, \\
\ee
where by the dots we denote corrections that are subleading in the derivative and loop expansions 
defining the effective theory. The K\"ahler metric computed from (\ref{stringk}) becomes diagonal 
and the whole Kahler manifold factorizes into the product of $n$ one-dimensional Kahler 
submanifolds. Also the only non-vanishing components of the Riemann 
tensor are the $n$ totally diagonal components ${R_{i \,\bar \jmath\, p\, \bar q} = 
R_{i} \,g_{i\,\bar \imath}^2 \, \delta_{i\, \bar \jmath p \bar q}}$ where $R_i={2}/{n_i}$. 
Recall now that when only chiral fields participate to supersymmetry breaking the flatness and 
stability conditions take the form (\ref{fs-1}), so in this particular case they just read:
\be\label{32}
\,\mbox{$\sum_i$}\, |f^i|^2 = 1\,,\hspace{1.5cm}\,\mbox{$\sum_i$}\, R_{i}\, |f^i|^4<  \frac 23 \,.
\ee
These relations represent a quadratic inequality in the variable $|f^i|^2$ subject to a linear constraint. 
This system of equations can be easily solved to get the condition $\sum_{i} R^{-1}_{i} 
> \frac32$, which translates into:
\be\label{con}
\hspace{4cm} \,\mbox{$\sum_i$}\, n_{i} > 3\,.
\ee
Also eqs.~(\ref{32}) constrain the values that the auxiliary fields {$|f_i|$} can take.

When a single modulus dominates the dynamics the condition (\ref{con}) implies $n>3$ (this result was 
already found in \cite{Brustein:2004xn} in a less direct way). For the universal dilaton $S$ we have 
$n_S = 1$ and therefore 
it does not fulfill the necessary condition (\ref{con}). This shows in a very clear way that 
just the dilaton modulus cannot lead to a viable situation \cite{nodildom} unless subleading corrections 
to its K\"ahler potential become large \cite{nonpertdil,yetmoredil}. 
We can therefore conclude that the scenario proposed in ref.~\cite{dilatondom}, in which 
the dilaton dominates supersymmetry breaking, can never be realized in a controllable way. 
On the other hand, the overall K\"ahler modulus $T$ has $n_T = 3$, and violates 
only marginally the necessary condition. In this case, subleading corrections to the K\"ahler potential 
are crucial. Recently some interesting cases where subleading corrections can help in achieving a 
satisfactory scenario based only on the $T$ field have been identified for example in 
\cite{nonpertvol,nonpertvolbis}.

In this case where the dynamics is dominated by 
just one field the K\"ahler potential of (\ref{stringk}) corresponds to a constant curvature manifold with
 $R = 2/n$ and it has a 
global symmetry associated to the Killing vector $X = i\, \xi$, which can be gauged as long as the 
superpotential is also gauge invariant. By doing so the potential would get a $D$-term contribution 
that should be taken into account in the analysis of stability, as was explained in the previous section. 
In such a situation the flatness condition in (\ref{fs-2}) can be solved by introducing 
an angle $\delta$ and parametrizing the rescaled auxiliary fields as $f = \cos \delta$ and 
$d = \sin \delta$. In terms of this angle the stability condition implies:
\be
n > \frac {3}{1 + 4\, \tan^6\delta} \,.
\label{c2}
\ee
From this expression, it is clear that it is always possible to 
satisfy the stability condition for a large enough value of $\tan\delta$.
Note in particular that eq.~(\ref{c2}) implies that when $n$ is substantially less than 
$3$, which is the critical value for stability in the absence of gauging, the contribution to 
supersymmetry breaking coming from the $D$ auxiliary field must be comparable to the 
one coming from the $F$ auxiliary field.

A final comment is in order regarding the issue of implementing the idea of uplifting with an uplifting
sector that breaks supersymmetry in a soft way. It is clear that such a sector will have to contain
some light degrees of freedom, providing also some non-vanishing $F$ and/or $D$ auxiliary field.
Models realizing an $F$-term uplifting are easy to construct. A basic precursor of such models 
was first constructed in \cite{lutysundrum}. 
More recently, a variety of other examples have been constructed, where the extra chiral multiplets 
have an O' Raifeartaigh like dynamics that is either genuinely postulated from the beginning 
\cite{Fup} or effectively derived from the dual description of a strongly coupled theory \cite{Fupiss} 
admitting a metastable supersymmetry breaking vacuum as in \cite{ISS}. Actually, a very simple 
and general class of such models can be constructed by using as uplifting sector any kind of sector 
breaking supersymmetry at a scale much lower than the Planck scale \cite{yo1}. Models realizing 
a $D$-term uplifting, on the other hand, are difficult to achieve. The natural idea of relying on some 
Fayet-Iliopoulos term \cite{BKQ} does not work, due to the already mentioned fact that such terms 
must generically be field-dependent in supergravity, so that the induced $D$ is actually 
proportional to the available charged $F$'s. It is then clear that there is an obstruction in getting 
$D$ much bigger than the $F$'s (see also \cite{cr}). 
Most importantly, if the only charged chiral multiplet in the model 
is the one of the would-be supersymmetric sector (which is supposed to have vanishing $F$) then 
also $D$ must vanish, implying that a vector multiplet cannot act alone as an uplifting sector
\cite{ChoiDup,DealwisDup}. This difference between $F$-term and $D$-term uplifting is, as 
was emphasized in the previous section,  
due to the basic fact that chiral multiplets can dominate supersymmetry breaking whereas vector 
multiplets cannot. 

Finally we would like to mention that the  flatness and stability conditions simplify not only for 
factorizable K\"ahler manifolds but also for some other classes of scalar manifolds that present  
a simple structure for the Riemann tensor. This is the case for example for K\"ahler potentials 
generating a scalar manifold of the form $G/H$ which arise for example in orbifold string
 models \cite{yo2,yo3}, and also for no-scale supergravities and 
Calabi-Yau string models \cite{yo4}.

\section{Conlcusions}

In this note we have reviewed the constraints that can be put on gauge invariant supergravity 
models from the requirement of the existence of a flat and metastable vacuum, following the 
results of \cite{yo1,yo2,yo3}. We have shown that in a general {${\cal N}=1$} supergravity theory 
with chiral and vector multiplets there are {strong necessary conditions} for the existence of 
phenomenologically viable vacua. Our results can be summarized as follows. These necessary conditions severely {constrain the geometry} of the scalar manifold as well as {the direction} of supersymmetry breaking and {the size} of the auxiliary fields. When supersymmetry breaking is 
dominated by the chiral multiplets the conditions restrict the {K\"ahler curvature}, whereas when 
also vector multiplets participate to supersymmetry breaking the net effect is to alleviate the 
constraints through a {lower effective curvature}. This is mainly due to the fact that the $D$-type 
auxiliary fields give a positive definite contribution to the scalar potential, on the contrary of the 
$F$-type auxiliary fields, which give an indefinite sign contribution. Nevertheless one should also 
take into account the fact that the local symmetries associated to the vector multiplets also 
restrict the allowed superpotentials. These results should be useful in discriminating 
more efficiently potentially viable models among those emerging, for instance, as low-energy
effective descriptions of string models. 

\begin{acknowledgement}
M.G.-R. would like to thank the organizers of the RTN meeting "Constituents, Fundamental Forces and 
Symmetries of the Universe" held in Valencia, 1-5 October 2007, for the opportunity to present this work. 
The work of C.A.S. has been supported by the Swiss National Science Foundation.
\end{acknowledgement}

\end{document}